\documentclass[aps,prl,reprint,floatfix,twocolumn,showpacs,superscriptaddress,longbibliography,nobalancelastpage]{revtex4-1}
\usepackage{graphics}
\usepackage{enumerate}
\usepackage{epsfig}
\usepackage{amsmath}
\usepackage{amssymb}
\usepackage[usenames]{color}
\usepackage[caption=false]{subfig}
\usepackage{hyperref}
\usepackage{float}

\begin{document}
\newcommand{\beq}{\begin{equation}}
\newcommand{\eeq}{\end{equation}}
\newcommand{\braket}[1]{\ensuremath{\left\langle #1 \right\rangle}}

\title{Jamming in Perspective}

\author{Varda F. Hagh}
\affiliation{Department of Physics, Arizona State University, Tempe, AZ 85287-1504, USA}

\author{Eric I. Corwin}
\affiliation{Department of Physics and Materials Science Institute, University of Oregon, Eugene, Oregon 97403, USA}

\author{Kenneth Stephenson}
\affiliation{Department of Mathematics, University of Tennessee, Knoxville, TN 37996, USA}

\author{M. F. Thorpe}
\affiliation{Department of Physics, Arizona State University, Tempe, AZ 85287-1504, USA}
\affiliation{Rudolf Peierls Centre for Theoretical Physics, University of Oxford, 1 Keble Rd, Oxford OX1 3NP, England}

\date{\today}

\begin{abstract}
Jamming occurs when objects like grains are packed tightly together (e.g. grain silos). It is highly cooperative and can lead to phenomena like earthquakes, traffic jams, etc. In this Letter we point out the paramount importance of the underlying contact network for jammed systems; the network must have one contact in excess of isostaticity {\it{and}} a finite bulk modulus.  Isostatic means that the number of degrees of freedom are exactly balanced by the number of constraints. This defines a large class of networks that can be constructed without the necessity of packing particles together compressively (either in the lab or computationally). One such construction, which we explore here, involves setting up the Delaunay triangulation of a Poisson disk sampling and then removing edges to maximize the bulk modulus, until the isostatic plus one point is reached. This construction works in any dimensions and here we give results in 2D where we also show how such networks can be transformed into a disk pack. 
\end{abstract}


\maketitle

Disordered packings of athermal frictionless particles are a standard model for studying the jamming transition in amorphous materials such as granular media~\cite{van2009jamming}, foams~\cite{bolton1990rigidity}, colloidal suspensions~\cite{puckett2011local}, and glasses~\cite{brito2006rigidity}. Every jammed system can be represented by a disordered spring network. To create this network, the center of mass of each particle is replaced with a vertex with an edge between two vertexes if their equivalent particles are in contact. The network embedding of a jammed system is isostatic {\it{plus one}}, meaning that the number of degrees of freedom ($dN$ where $d$ is the dimension and $N$ is the number of vertexes) and constraints ($N_{e}$ that is the number of edges) are balanced in a way that there is exactly one state of self stress in the system. This extra {\it{plus one}} is necessary for mechanical stability and a finite bulk modulus~\cite{thorpe2009networks,guyon1990non}. This then becomes a combinatoric rather than a geometry problem as only the network topology is involved; assuming the network is generic (no symmetry) which is the case in disordered networks, glasses etc. The Maxwell count for an isostatic system, which has a {\it{periodic super cell}}, is such that the number of floppy modes, $F$, are exactly zero, so 

\begin{equation}
\label{eqn:count}
F = dN - N_{e} -d =0
\end{equation}
with the dimension $d = 2$ in this Letter. The last term is to make sure that the $d$ macroscopic translations are properly accounted for. 

We use the pebble game~\cite{jacobs1995generic,jacobs1997algorithm} (a numerical algorithm based on Laman's theorem~\cite{laman1970graphs}) in $2$D to determine the rigid region decomposition of the network. For jammed systems at the isostatic point, the system is isostatic everywhere, with no stressed edges. We refer to this as {\it{locally isostatic}}~\cite{theran2015anchored}. This is a stricter requirement than just applying Eq.~(\ref{eqn:count}) once globally, as it requires that all subgraphs are also isostatic.  Clearly just applying (\ref{eqn:count}) globally could give locally stressed regions balanced by other regions containing floppy modes and hinges, as happens in rigidity percolation~\cite{ellenbroek2015rigidity}.  

Traditional computational methods available to create jammed packings, usually with disks or spheres, include some mixture of molecular dynamics, event driven dynamics, and energy minimization schemes \cite{ohern_random_2002, donev_pair_2005, liu2010jamming, torquato_robust_2010, goodrich_finite-size_2012, lerner_low-energy_2013}.  The new method introduced here, produces a jammed network with precisely one state of self stress and expands the set of what was previously accepted as jammed. To be precise, we define a jammed network as being {\it{isostatic plus one excess contact}} and having a {\it{finite bulk modulus}}. By finite we mean $O(1)$ and not $O(1/N)$ which will go to zero as the number of vertexes $N$ tends to infinity. Such a network has the consequence that when one edge is removed, the network is locally isostatic. With this definition, we are now free to adopt any construction method that will achieve this. There is the traditional method which packs particles together by compression and a new method described here. Other definitions of jammed systems are available (see Theorem 1 in ~\cite{connelly2017isostatic}) but we have found the above to be the most useful in practice. 

\begin{figure*}[t]
\subfloat[]{\includegraphics[width=6cm]{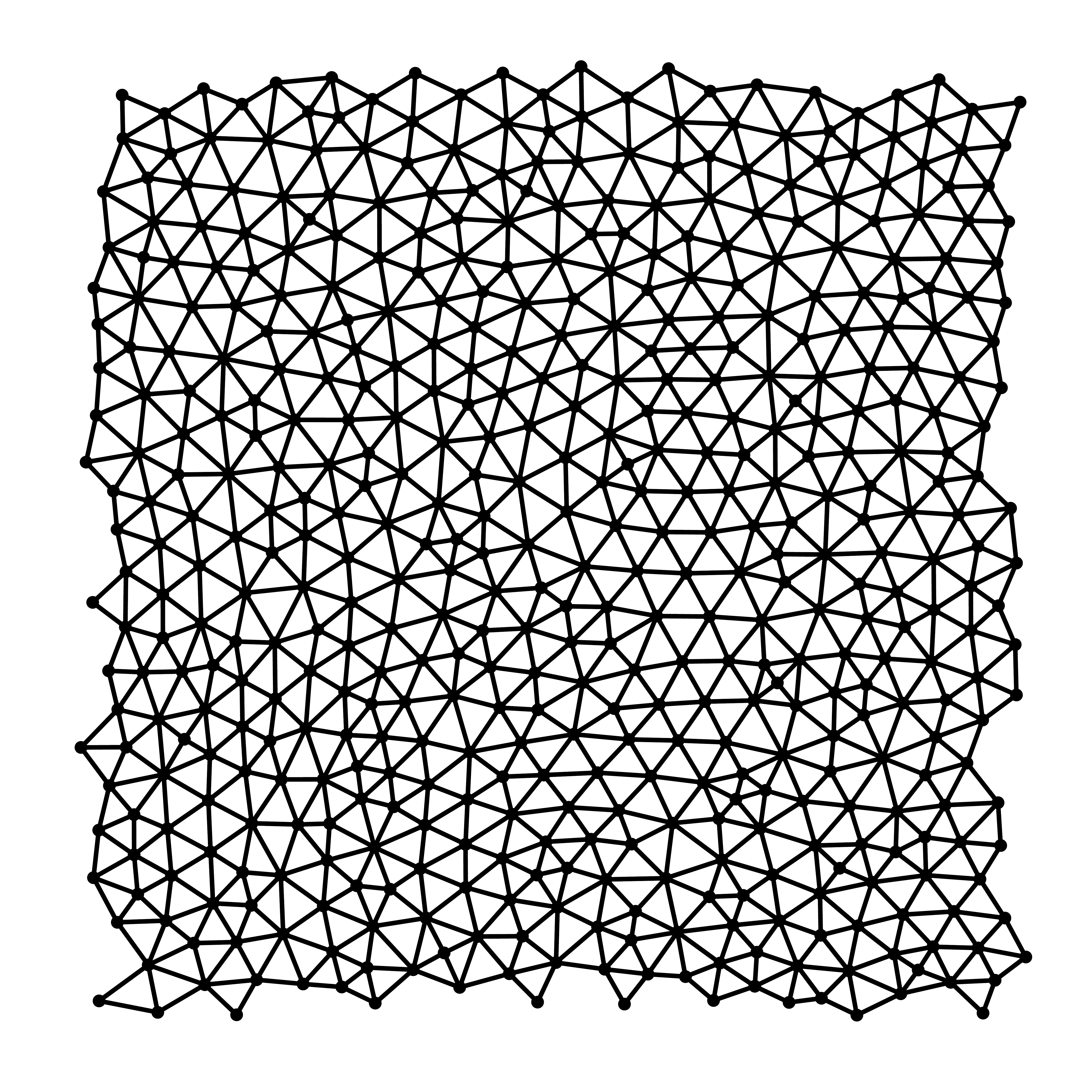}}
\subfloat[]{\includegraphics[width=6cm]{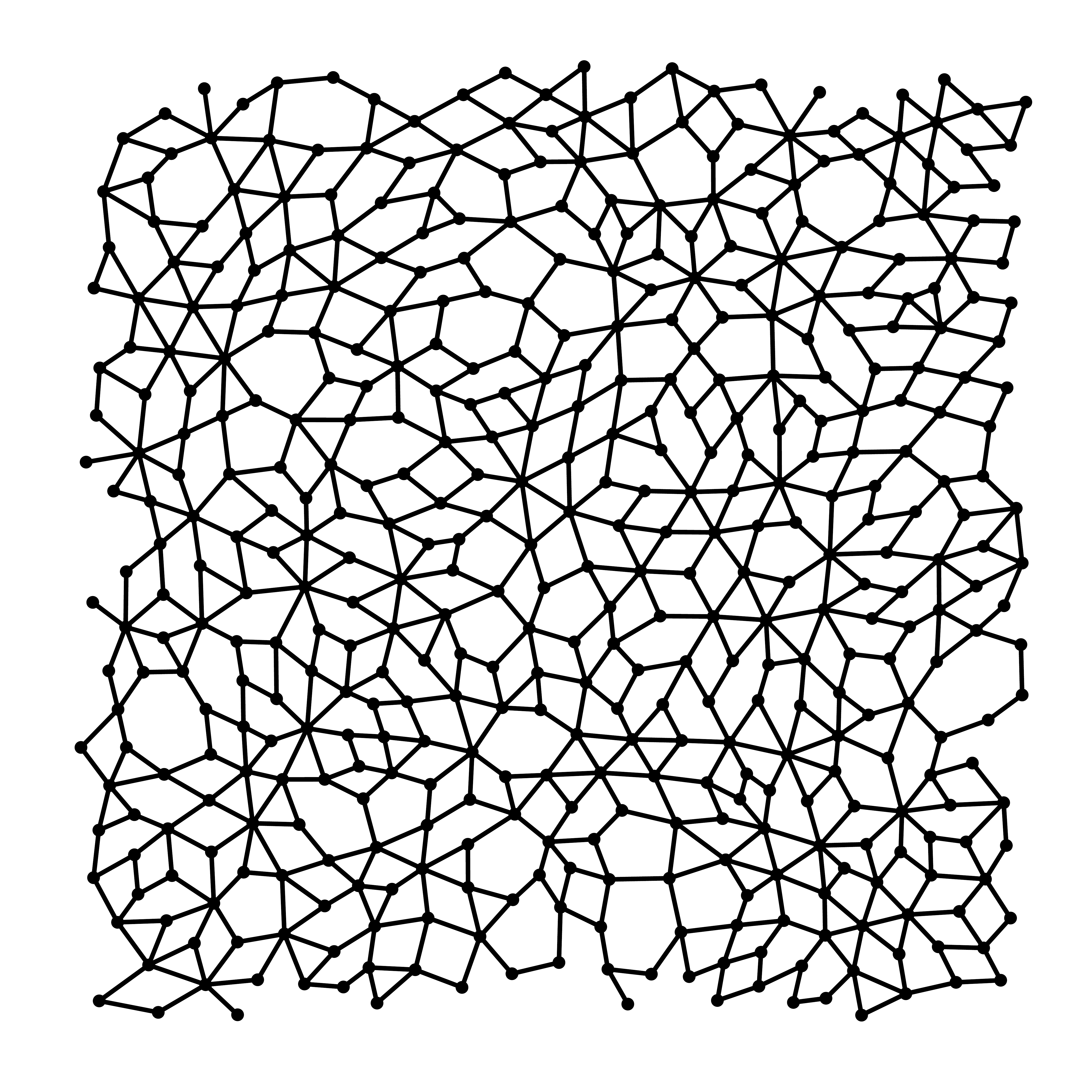}} 
\subfloat[]{\includegraphics[width=6cm]{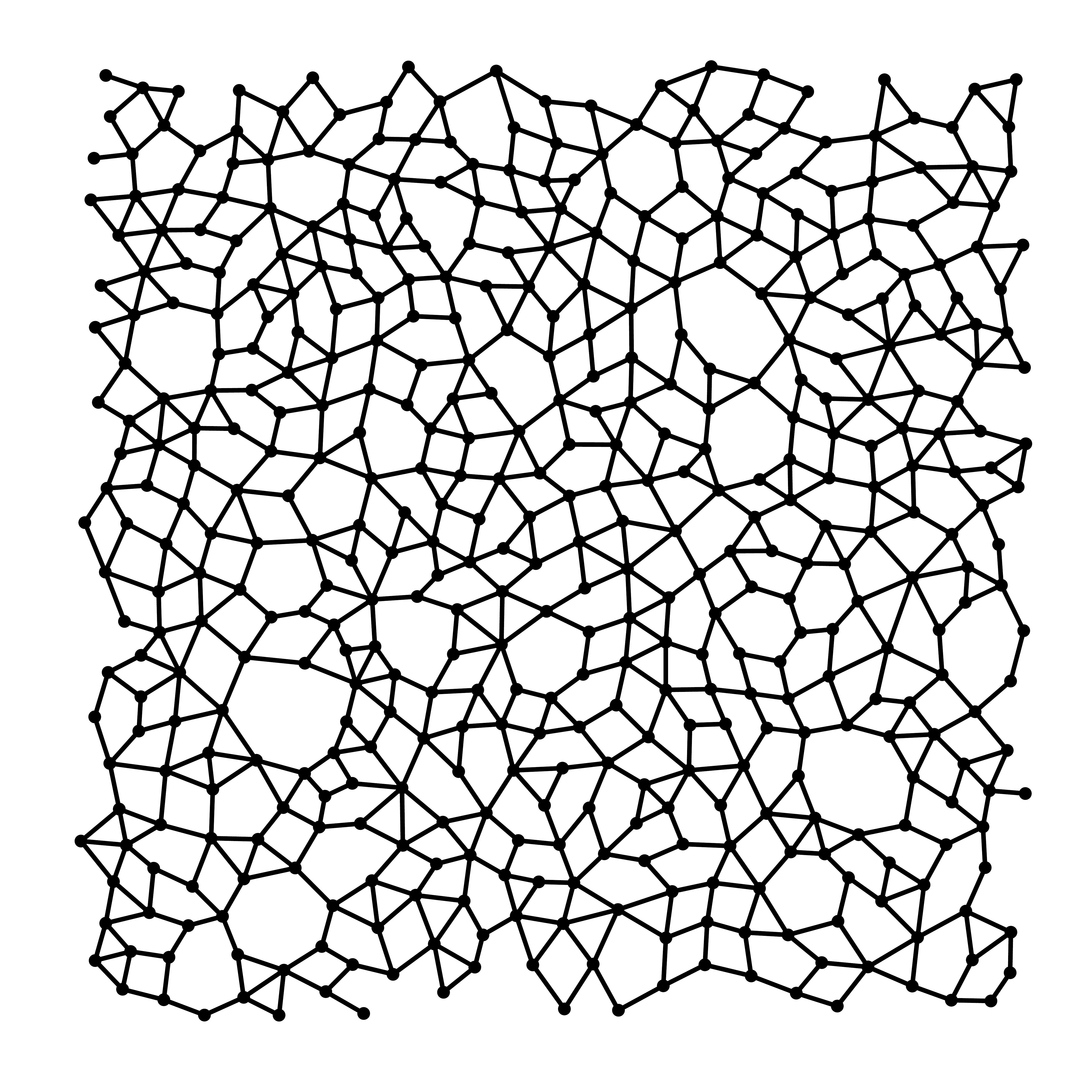}}
\caption{a) Delaunay triangulation of a Poisson disk sampling with 512 points. b) The same network at the isostatic plus one, after pruning edges that minimally reduce the bulk modulus and removing the rattlers. c) The network representation of a polydisperse jammed pack, formed by compressing disks, with approximately same number of vertexes as in part (b).}
\label{fig:Delaunay}
\end{figure*}

The new approach uses an  algorithm that allows for precise control over the number of contacts in excess of isostaticity \cite{charbonneau_universal_2012, morse_geometric_2014, charbonneau_jamming_2015}. We focus on the network as being fundamental to the jammed state and show that in two dimensions, the network can always be replaced by a disk pack, as well as vice-versa. Often it is useful to add a single additional edge (or contact) to create a single state of self stress and we will refer to this as {\it{isostatic plus one}}.  We note that this is often referred to confusingly as isostatic in the literature and we strongly discourage this usage. These systems are delicately balanced and a single edge present at isostaic plus one does make a global difference at the isostatic point; no matter how large the system. 

For a non-crystalline system to be jammed it is necessary but not sufficient for it to be isostatic plus one. An additional degree of cooperativity needs to be introduced by demanding that the bulk modulus drops from finite to zero as a single edge is removed in going from isostatic plus one to the isostatic state. A locally isostatic network can be easily achieved by randomly removing stressed edges from a highly over-constrained network, but the resulting network will not necessarily have a finite bulk modulus at isostatic plus one~\cite{ellenbroek2015rigidity}. Therefore the finiteness of bulk modulus does \textit{not} follow from the system being locally isostatic when an edge is removed. 
A convenient way to characterize the extreme cooperativity of jammed networks is through two indexes $s$ and $h$, where $s$ measures the fraction of stressed edges, when any one additional edge is added to an isostatic network, and $h$ measures the fraction of hinged vertexes when any one edge is removed. This comes entirely from the static properties, using the pebble game, and is a very convenient way to establish the marginality of jammed networks without getting into the details of low frequency dynamics~\cite{wyart2005geometric,silbert2005vibrations} which is discussed in detail in the Supplemental Material. If rattlers are removed, both locally isostatic and jammed networks can have $s=1$ and $h=1$ ~\cite{ellenbroek2015rigidity, NoteX}, so this cannot be used to distinguish between them. Hence we need to include in the definition of jammed states that the bulk modulus is finite at isostatic plus one.

The new method to generate polydisperse jammed packs at zero temperature does not require exploring the entire energy landscape to bring the system into zero internal energy and isostaticity. Instead, it builds the system within a single local energy minimum. We try to keep cavities to a minimum so all packing fractions are within the range $ 0.77 < \phi < 0.82$ after removing the rattlers.

\begin{figure}[!tb]
\includegraphics[width=8cm]{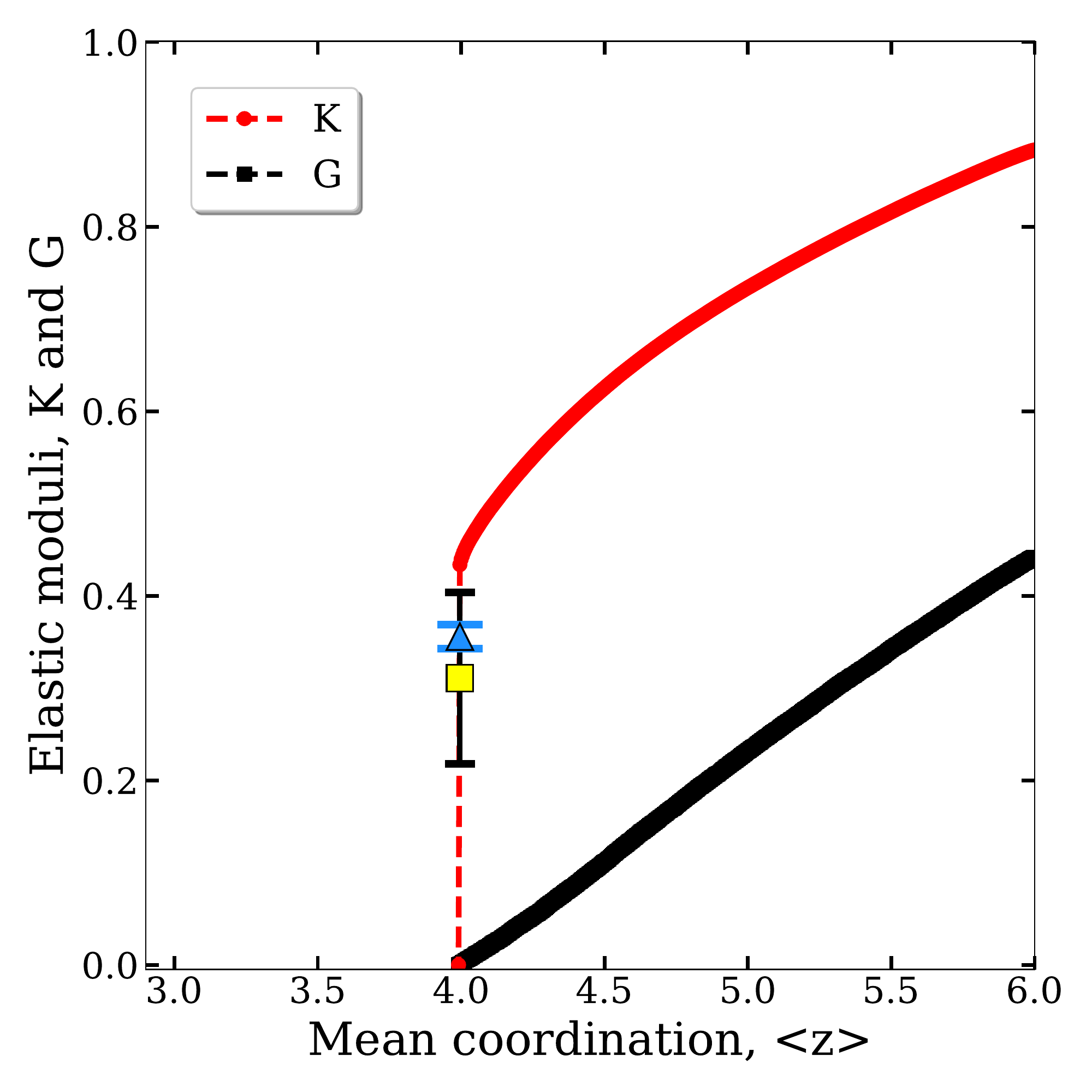}
\caption{(color online). The ensemble averaged bulk $K$ (red) and shear $G$ (black) elastic moduli of $100$ samples with $512$ vertexes as the edges are removed from mean coordination $\braket{z} = 6$ down to $\braket{z} \simeq 4$. The yellow square, with a wide spread, shows the average of bulk moduli for $100$ samples generated by {\tt CirclePack}. The blue triangle, with a tighter spread, shows the average of bulk moduli at isostatic plus one for $100$ samples generated by conventional jamming algorithms. The jammed systems have the same disk size distribution as circle packs.}
\label{fig:bulk}
\end{figure}

This new method is based on a pruning algorithm that is used to manipulate and control the elastic properties of disordered harmonic spring networks~\cite{goodrich2015principle}. These disordered networks are usually created by minimizing the energy of $N$ repulsive frictionless particles in a periodic box and stopping at a coordination that is slightly above jamming transition point. Therefore they already have encoded in them the properties of jamming and should not be thought of as generic networks. By contrast, in this work we generate the initial networks \textit{de novo and far from jamming}, using computational geometry only. The disordered jamming-like networks are then created by performing a simple set of steps. A summary of the procedure is presented below:

\begin{itemize}
\item We start by generating $N$ points in a box with periodic boundary conditions that are distributed by Poisson disk sampling~\cite{bridson2007fast,dunbar2006spatial}. We then find the Delaunay triangulation of these points~\cite{lee1980two}.

\item To make the triangles more regular, we move each vertex to the centroid of the polygon formed by its nearest neighbors, iteratively, until every vertex is at the centroid of its neighbors. An example of such generated samples is shown in Figure~\ref{fig:Delaunay}-a. This geometrically generated network is highly over-constrained and far from isostatic (with a mean coordination of $\braket{z}=2N_e/N=6$), therefore we need to remove $N_r$ redundant edges to push it down to the isostatic plus one point as desired.

\item There are $\binom{N_{e}}{N_r}$ ways to prune these $N_r$ redundant edges from the network. It is well known \cite{hexner2018role,hexner2017linking} that the contribution of a removed edge to the bulk modulus is largely independent of its contribution to the shear modulus, although these moduli cannot increase by removing an edge (~\cite{lord1945theory}, pp. 110-111).  Since jammed packs maintain a finite bulk modulus while the ratio of shear ($G$) and bulk ($K$) moduli vanishes at jamming point~\cite{o2003jamming}, at each step we find and remove the edge that maximizes the bulk modulus of the remaining network. Maximizing the bulk modulus is not strictly necessary as similar results can be obtained if we remove an edge randomly from the top 20\% of edges that have minimal contribution to the changes in bulk modulus. 

\item We repeat the process, until we arrive at isostatic plus one where $\braket{z} \simeq 4$. The resulting network has a finite bulk modulus and is shown in Figure~\ref{fig:Delaunay}-b. Figure~\ref{fig:bulk} shows how the bulk and shear elastic moduli of the network change as the edges are pruned. The behavior of the shear modulus is reminiscent of random rigidity percolation models~\cite{ellenbroek2015rigidity} as well as jamming.

\end{itemize}

At this point we have a spring network that is identical to the network representation of a jammed pack (an example is shown in Figure~\ref{fig:Delaunay}-c) in all the following aspects (none of which holds for a percolating rigid network at the marginal point):

\begin{enumerate}[I]
\item The network has one excess contact past mathematical isostaticity (isostatic plus one),
\item The bulk modulus of the network is finite and $O(1)$,
\item The ratio of shear and bulk elastic moduli ($G/K$) scales as $\Delta z = \braket{z} - z_J$ where $z_J$ is the mean coordination at the marginal point,
\item It is marginal, as both its $s$ and $h$ indexes are equal to $1$~\cite{NoteX} and its density of states for low excitation frequencies is akin to that of a jammed system as is shown in the Supplemental Material,
\item It is stable as revealed by the study of its dynamical matrix. All of the eigenvalues are positive (except for the two trivial translational eigenvectors whose eigenvalues are zero).  
\item 100\% of the forces along the edges in the network are positive definite and their distribution exhibits a scaling behavior similar to jamming [see the Supplemental Material]. This is very different from percolating networks at the critical point where the fraction of compressive forces is about 50\%.
\end{enumerate}

\begin{figure}[!tb]
\subfloat[]{\includegraphics[width=7cm]{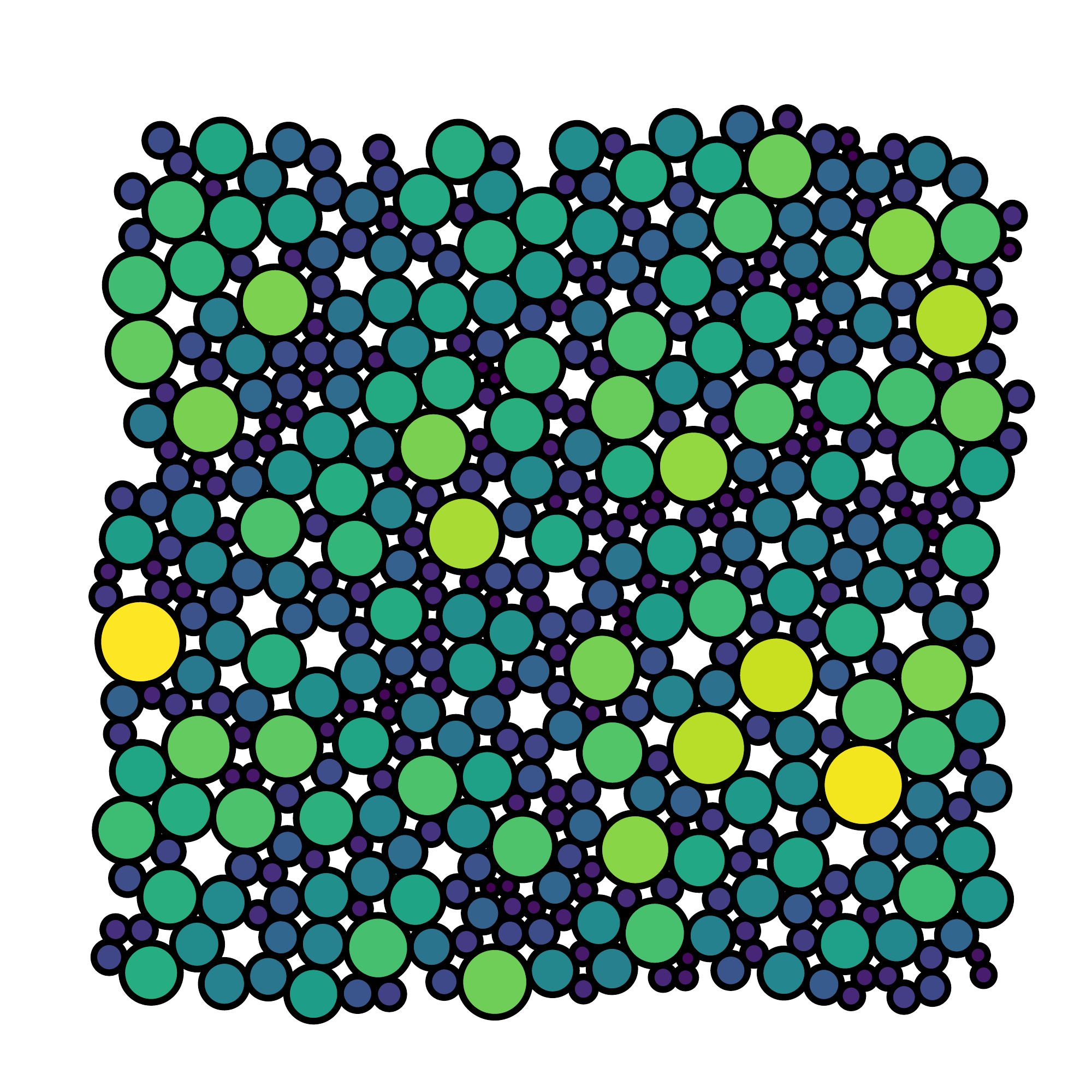}}
\vspace{-0.5cm}
\subfloat[]{\includegraphics[width=7cm]{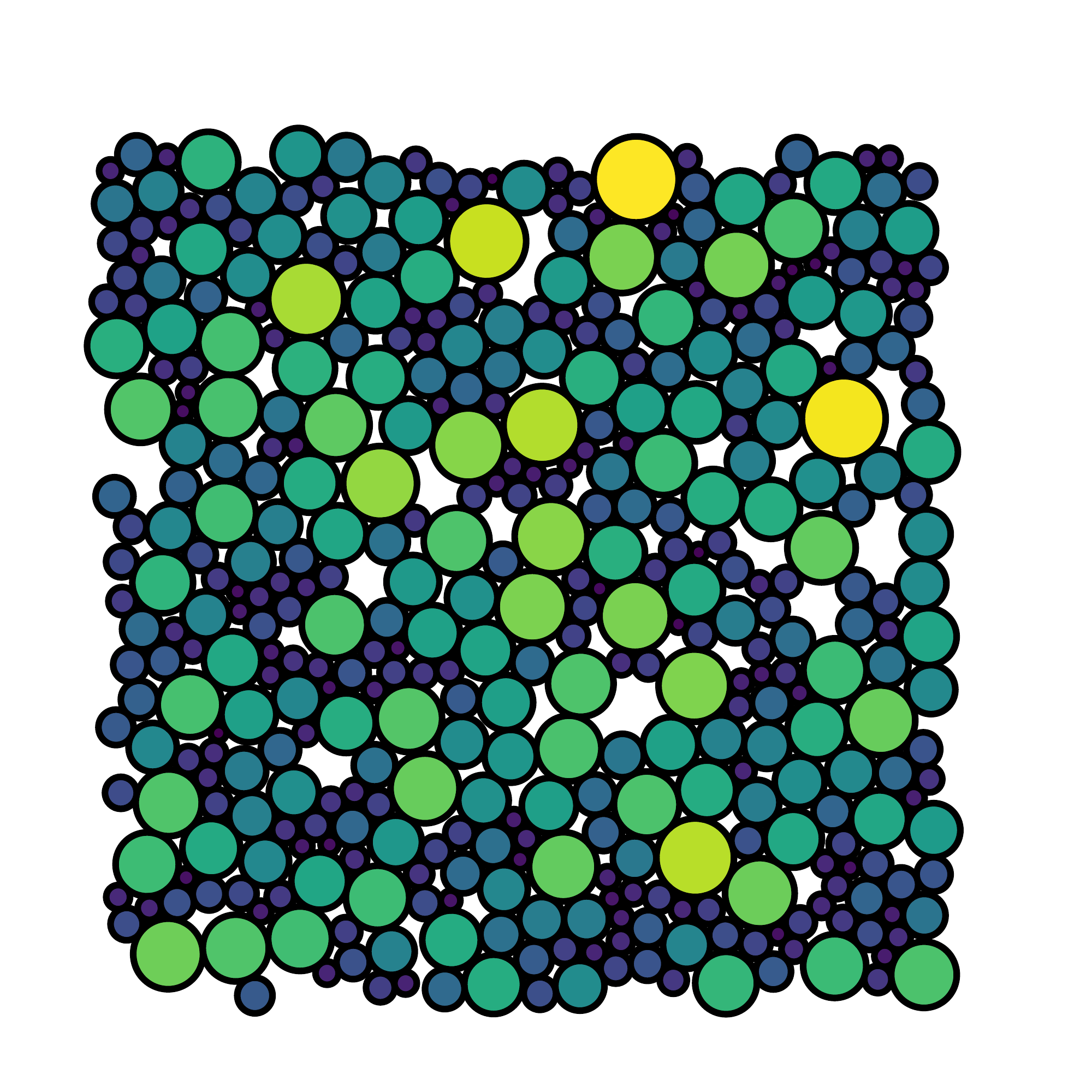}}
\caption{(color online) a) Packing generated by pruning algorithm and {\tt CirclePack} b) Rattler free packing generated by standard algorithms.}
\label{fig:circle_packs}
\end{figure}

This network can now be mapped into a disk packing~\cite{lopez2013jamming}. We locate disks for a given periodic network using methods of circle packing, a topic introduced by William Thurston, \cite{wT-t,wT85}; the standard reference is \cite{kS05}, see in particular Chapter~9. A {\it circle packing} (or disk packing) is a configuration of circles satisfying a prescribed pattern of tangencies. In our setting, prescribed tangencies are those of the given network, which is treated as a graph on a topological torus. Computations are carried out in the software {\tt CirclePack}, \cite{kS92}. They require a triangulation, so a single auxiliary vertex is temporarily added to each complementary cell of the network. For the resulting triangulation, circle packing theory (see \cite{BSt90} and \cite{kS05}[Prop~9.1]) guarantees the existence of a geometric torus and an associated circle packing on that torus.  {\tt CirclePack} computes disk radii and lays the disks out as a periodic circle packing in the plane.  While the result of {\tt CirclePack} is unique up to scaling and rigid motions, there are many such packings that could satisfy the constraints of the original network. Discarding the disks for the auxiliary vertexes leaves a circle packing with locations and radii for the vertexes of the original network, as in Figure~\ref{fig:circle_packs}-a.

{\tt CirclePack} changes the geometrical configuration of vertexes. However, the connectivity of the system does not change and the bulk modulus remains finite after this transformation with a standard deviation of $s = 0.09$ for the samples studied here, as can be seen in Figure~\ref{fig:bulk}. 

The generated circle packing holds all but one of the properties of the pruned networks discussed above. It is at isostatic plus one, has a finite bulk modulus of $O(1)$ and a vanishingly small shear modulus of $O(1/N)$. It is also marginal with $s=h=1$, and stable which means it would not change for a small enough compress-decompress protocol. The difference is that not all the forces in the system (although a majority of 72\% to 99\% of them in the samples studied here) are necessarily positive definite (item \MakeUppercase{\romannumeral 6} above). This comes as a result of our non-unique mapping from the network to the disk packing.

Every circle packing has a distribution of radii that can be assigned to particles in a standard molecular dynamics simulation to generate a polydisperse $2$D disk packing that can be compared to the packing generated by the newly introduced algorithm.
In this approach, we first scale the radii of particles to achieve a starting packing fraction well above the jamming transition; typically packing fraction $\phi_J \simeq 0.85$ for disks.  Particles interact through a standard contact harmonic potential.  The system is minimized to its inherent structure at this initial density using a quad-precision GPU implementation of the FIRE algorithm \cite{bitzek_structural_2006, charbonneau_jamming_2015}.  Configurations at a desired excess number of contacts can be achieved by exploiting the scaling of total energy $U \propto (\phi - \phi_J)^2$, where $\phi_J$ is the isostatic jamming density.  The system is successively brought to lower energies and thus lower numbers of excess contacts by rescaling the radii and re-minimizing.  The re-scalings are chosen to achieve approximately 10 steps per decade of $\phi - \phi_J$.  This process continues until the number of excess contacts is reduced to the desired value.  At each density the number of excess contacts is calculated on the rigid core of the system by first removing {\it{rattler}} particles lacking at least $d+1$ non-cohemispheric contacts. The blue triangle in Figure~\ref{fig:bulk} shows the average bulk modulus of $100$ samples generated by this method. The standard deviation is in order of $s = 0.01$, which is smaller than the standard deviation obtained from results of {\tt CirclePack}. 

There are measurables that are not universal - like the density, pair distribution function, etc.  These vary widely for conventional jammed packs as well as in the jammed systems here, depending largely upon the number of rattlers, the size of convex cavities that are present, and the protocol that is being used to generate the jammed packs. For instance, the average packing fraction of $100$ test samples generated by {\tt CirclePack} is $\phi \simeq 0.77$ which is lower than that of samples generated by our standard algorithm where $\phi \simeq 0.82$ after removing the rattlers. We emphasize again that the circle packing construction used here is not unique and does not create packings with all positive definite forces. This then explains the lower density as it is well known that attractive interactions (or indeed frictional interactions) allow one to create critically jammed packings at significantly lower densities. The precise ways the disks of various radii are located is also not a crucial issue and can vary from well mixed to some clustering. Figure~\ref{fig:circle_packs} shows the comparison of two samples with 512 particles.

In this Letter, we have shown that the essence of the jamming transition is the underlying network involved at the isostatic plus one point. But another ingredient is required - that the bulk modulus goes from a finite value to zero as one constraint is removed to take the network from isostatic plus one to isostatic. This not only clarifies the nature of the jamming transition, but shows that conventionally jammed networks (formed by compacting particles together) are part of a larger group of networks controlled by topology with the added cooperative geometric ingredient that the bulk modulus remains finite. Such cooperativity is essential to make the network jammed, and much more restrictive than merely being isostatic. We have also demonstrated that all of the interesting macroscopic properties of jammed matter derive from the marginality of the system and its bulk mechanical properties. As such, both our generated networks and their equivalent circle packings behave as properly jammed systems for all bulk interrogations. However, the microscopic properties of jamming are only satisfied by the pruned networks and not the circle packs. This is because the force distributions in pruned networks and jamming follow similar scaling laws, whereas the circle packings fail to do so since forces are not positive everywhere. We note finally that in all the networks discussed in this Letter, the shear modulus goes from $O(1/N)$ at isostatic plus one, to zero at isostatic. The ideas in this Letter generalize easily to any dimensions, but the final step of going from a network to a hypersphere pack is only possible in $2$D. 

We acknowledge discussions with Wouter Ellenbroek and Louis Theran on the properties of isostatic disordered networks. The work at Arizona State University is supported by the National Science Foundation under grant DMS 1564468. EIC is supported by the NSF under Career Grant No. DMR-1255370 and a grant from the Simons Foundation No. 454939. This work used the Extreme Science and Engineering Discovery Environment (XSEDE), which is supported by National Science Foundation grant number ACI-1548562.  Specifically, it used the Bridges system, which is supported by NSF award number ACI-1445606, at the Pittsburgh Supercomputing Center (PSC). This work also used the University of Oregon high performance computer, Talapas. We gratefully acknowledge the support of NVIDIA Corporation with the donation of a Titan X Pascal GPU used in part for this research.
\bibliography{jamming}

\begin{thebibliography}{10}
\bibitem{NoteX}
When pruning a spring network, if we do not remove the rattlers that appear in
  the form of vertexes with coordination number $z = 2$, the $s$ and $h$
  indexes will be slightly smaller than 1.
\end{thebibliography}

\clearpage
\onecolumngrid
\section{Supplemental Material}
\subsection{Vibrational Modes}
Here we look into the density of states (DOS) in the pruned network constructions and their equivalent circle packs and compare the results to physically jammed systems. First, we study the evolution of DOS in the disordered networks as they are pruned from $\braket{z} = 6$ to $\braket{z} \approx 4$. 
For a $2$D spring network of area $A$, the number of allowed wave modes between wave numbers $0$ and $q$ is~\cite{kittel1996introduction}:

\begin{equation}
\label{eqn:DOS_count}
n(q) = \frac{A}{(2\pi)^2}\, \pi q^2
\end{equation}

We assume the vibrational frequencies are low enough for the dispersion relation to be almost linear for both longitudinal ($L$) and transverse ($T$) acoustic modes:

\begin{equation}
\label{eqn:dispersion}
q = \frac{\omega}{v_{\alpha}}
\end{equation}
where $\alpha = T, L$. This means the number of vibrational modes $n(\omega)$ is quadratic in frequency which leads to the following form for density of states:
 
\begin{equation}
\label{eqn:DOS}
\mathcal{D}(\omega) = \frac{d n(\omega)}{d\omega} = \frac{A}{2\pi v_{\alpha}^2}\, \omega
\end{equation}

\begin{figure}[h]
\centering
\includegraphics[width=10cm]{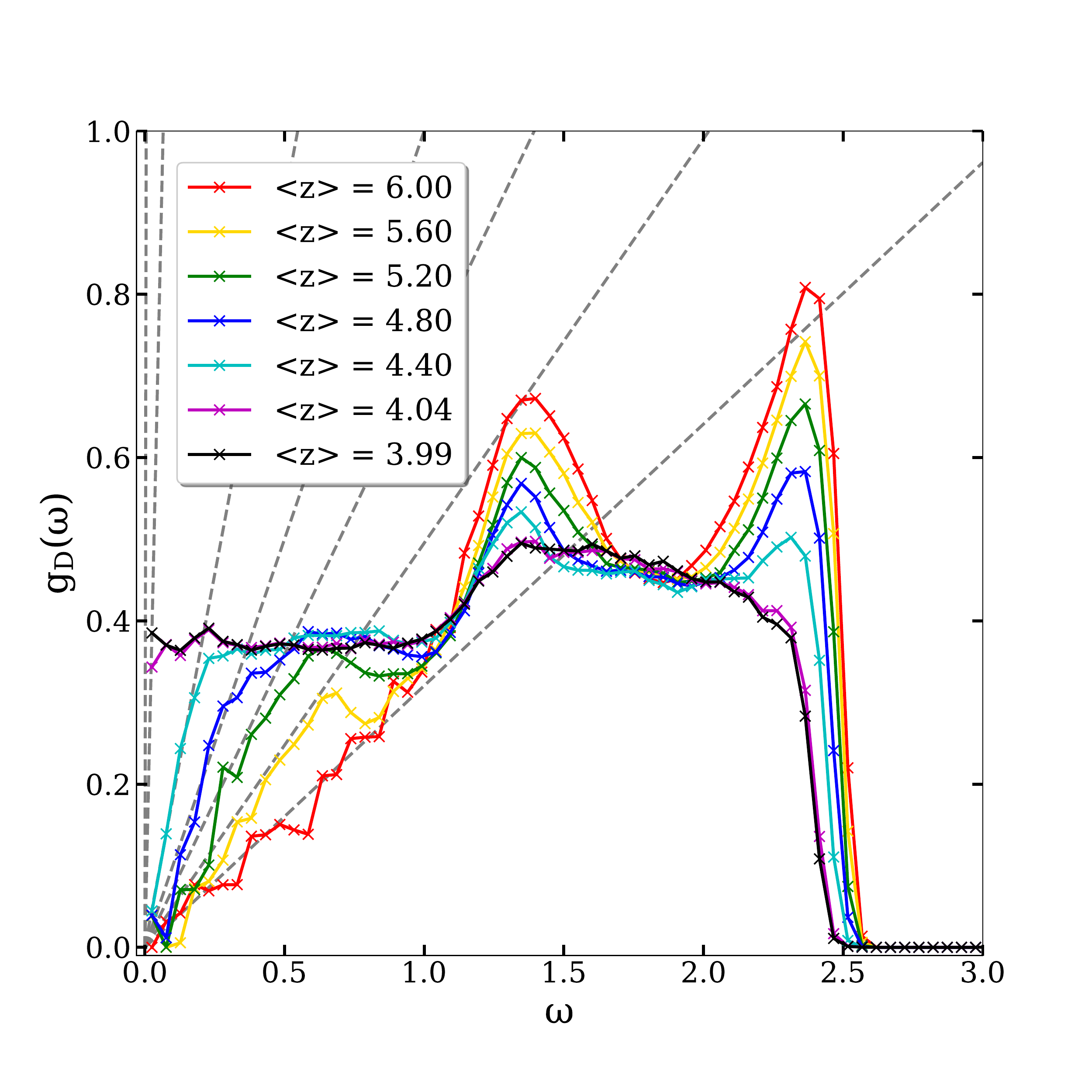}
\caption{(color online) The evolution of probability density function for acoustic modes in disordered spring networks as the bonds are pruned from $\braket{z} = 6$ down to $\braket{z} \approx 4$ (isostatic plus one) while keeping the bulk modulus finite. The dashed lines display Eq.~(\ref{eqn:PDF_DOS}) for the average elastic moduli associated with each value of $\braket{z}$ shown on the colored curves. The results are ensemble averaged over $100$ samples, each with $512$ vertexes.}
\label{fig:DOS-full_z}
\end{figure}

On the other hand, the longitudinal and transverse sound velocities are related to the bulk ($K$) and shear ($G$) moduli of a $2$D spring network in the following form:
\begin{align}
\label{eqn:long-trans}
v_L &= \sqrt[]{\frac{G+K}{\rho}} \nonumber\\
v_T &= \sqrt[]{\frac{G}{\rho}}
\end{align}

where $\rho = N/A$ is the mass density. Here the mass density is equal to the number density of the system since all vertexes have unit mass. 
By inserting Eq.~(\ref{eqn:long-trans}) into Eq.~(\ref{eqn:DOS}) and using the normalization $g_{\mathcal{D}}(\omega) = \mathcal{D}(\omega)/N$ so that $\int_{}^{} g_{\mathcal{D}}(\omega)\, d\omega = 1$, we can write the probability distribution function of the vibrational modes in terms of the elastic moduli of the system~\cite{zhang2017experimental}:

\begin{equation}
\label{eqn:PDF_DOS}
g_{\mathcal{D}}(\omega) = \frac{\omega}{2\pi} (\frac{1}{G} + \frac{1}{G+K}) 
\end{equation}

The linearity of $g_{\mathcal{D}}(\omega)$ versus $\omega$ is the Debye-like low frequency  behavior that is expected to be seen in any material with non-zero values of sound velocities. This is observed for networks far from marginality in the lower left corner of Figure~\ref{fig:DOS-full_z}. When the edges with smallest contribution to the bulk modulus are removed from a fully triangulated disordered spring network, the shear modulus approaches zero almost linearly, while the bulk modulus remains finite. Therefore the first term on the RHS of Eq.~(\ref{eqn:PDF_DOS}) diverges and the density of states becomes flat near the transition point which is a characteristic of the vibrational modes in disordered systems at their marginal transition point~\cite{van2009jamming, charbonneau2016universal,zhang2017experimental}.

\begin{figure}[!h]
\subfloat[]{\includegraphics[width=7cm]{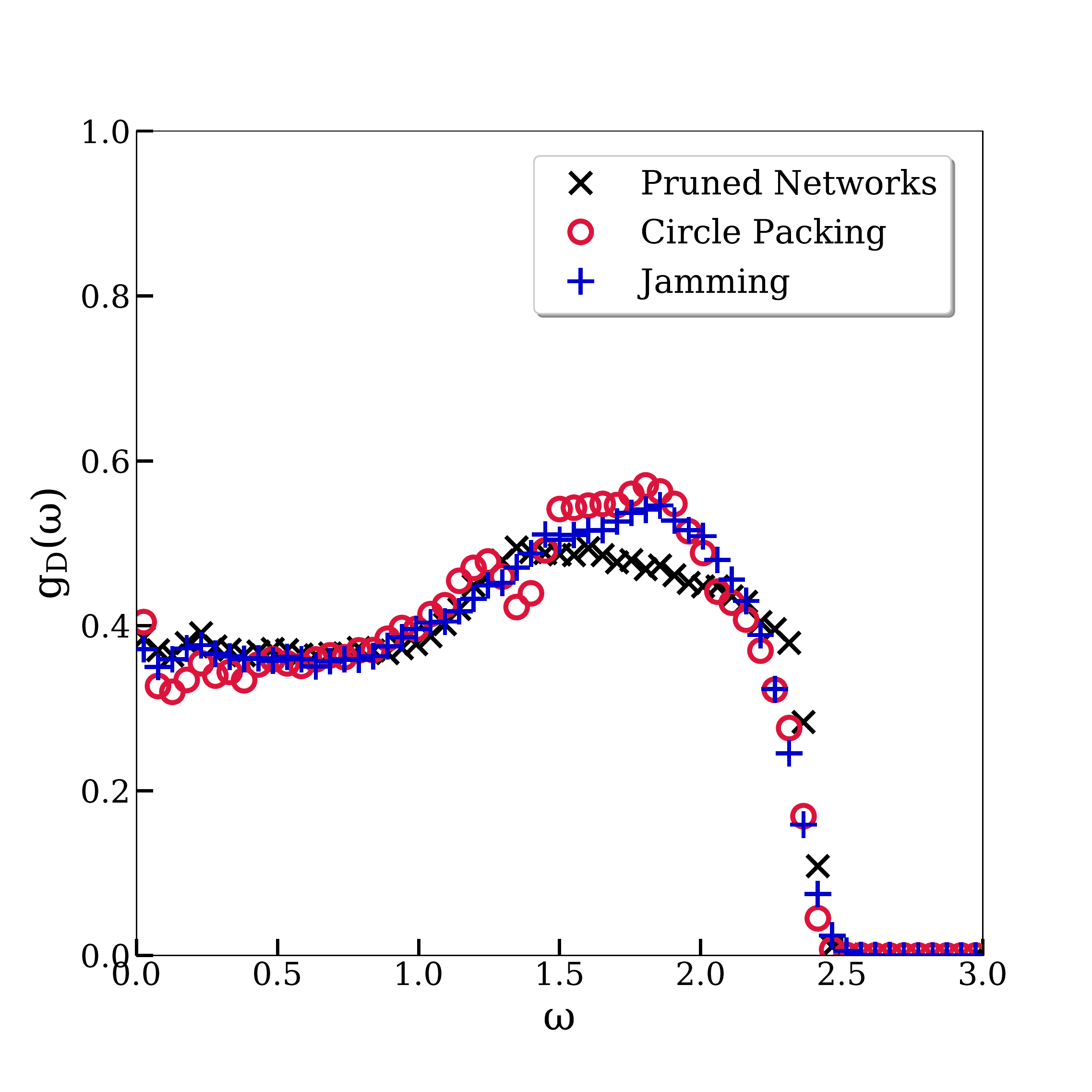}}\quad \quad 
\subfloat[]{\includegraphics[width=7cm]{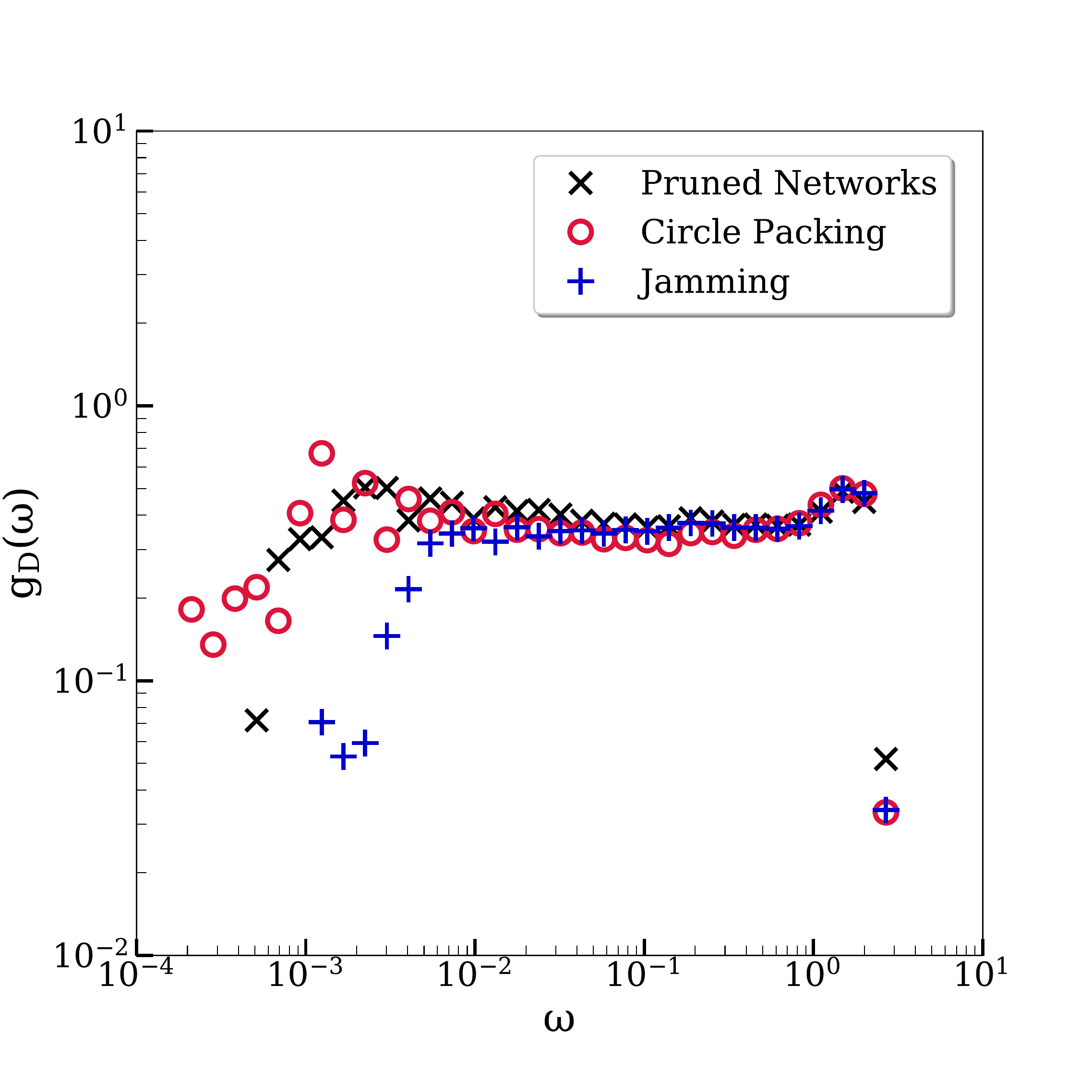}}
\caption{(color online) a) The probability density function for vibrational modes in $2$D pruned networks (blue), their equivalent circle packs (red) and jammed systems (black) in linear scale. b) The plot of part (a) in logarithmic scale.}
\label{fig:DOS_all_3}
\end{figure}

Figure~\ref{fig:DOS_all_3} shows the plots of $g_{\mathcal{D}}(\omega)$ for three types of systems studied in the Letter: the pruned networks at isostatic plus one, their equivalent circle packings, and the jammed systems generated by using the size distribution of circle packs both in linear and logarithmic scale. The marginality of all these systems is evident by their flat density of states at low frequencies.

\subsection{Distribution of Forces}

\begin{figure}[!t]
\centering
\includegraphics[width=8cm]{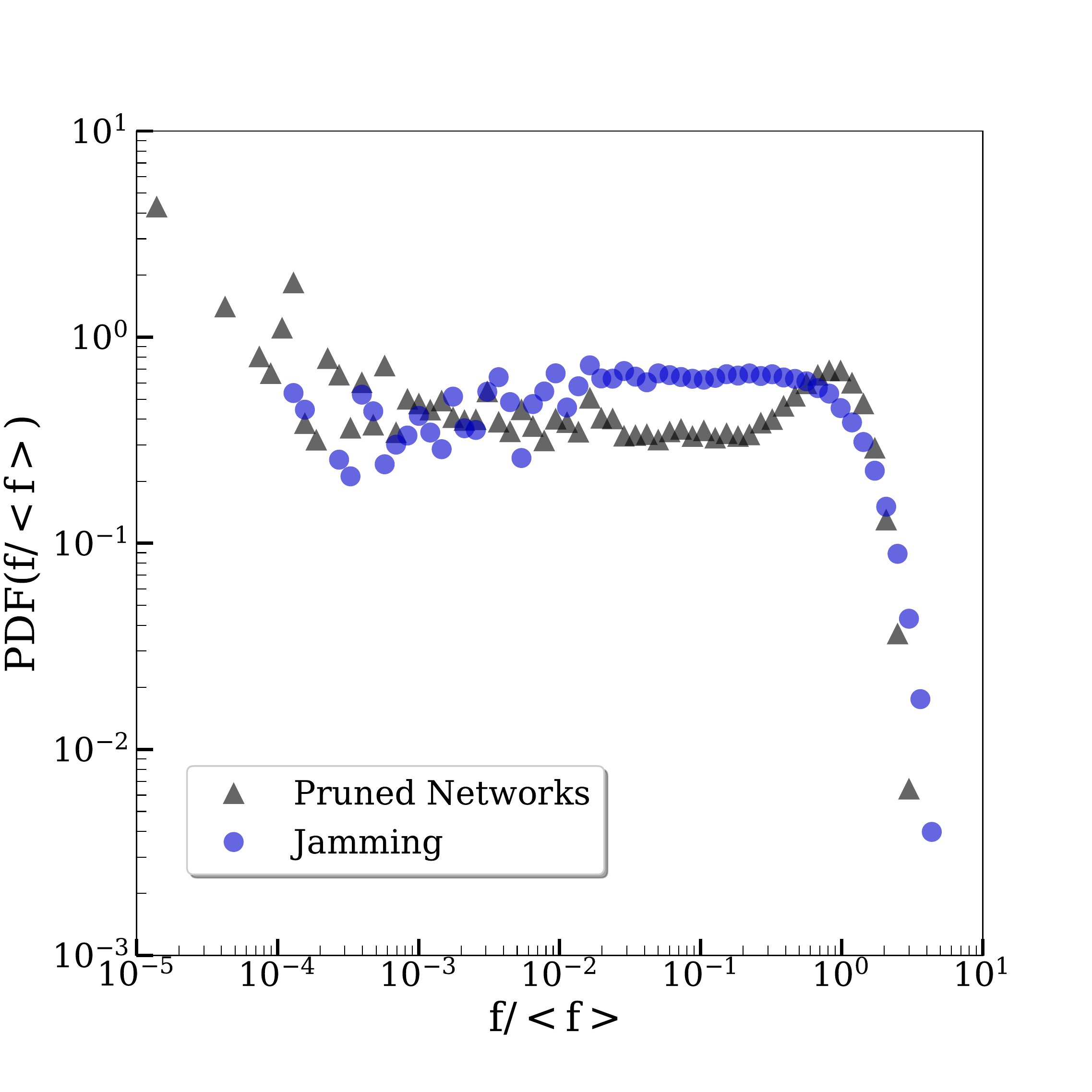}
\caption{(color online) The probability distribution function of forces for pruned networks (gray triangles) and jammed systems (blue circles) at isostatic plus one.  Both exhibit a nearly constant distribution of forces for small forces.}
\label{fig:PDForces}
\end{figure}

\begin{figure}[!t]
\centering
\includegraphics[width=8cm]{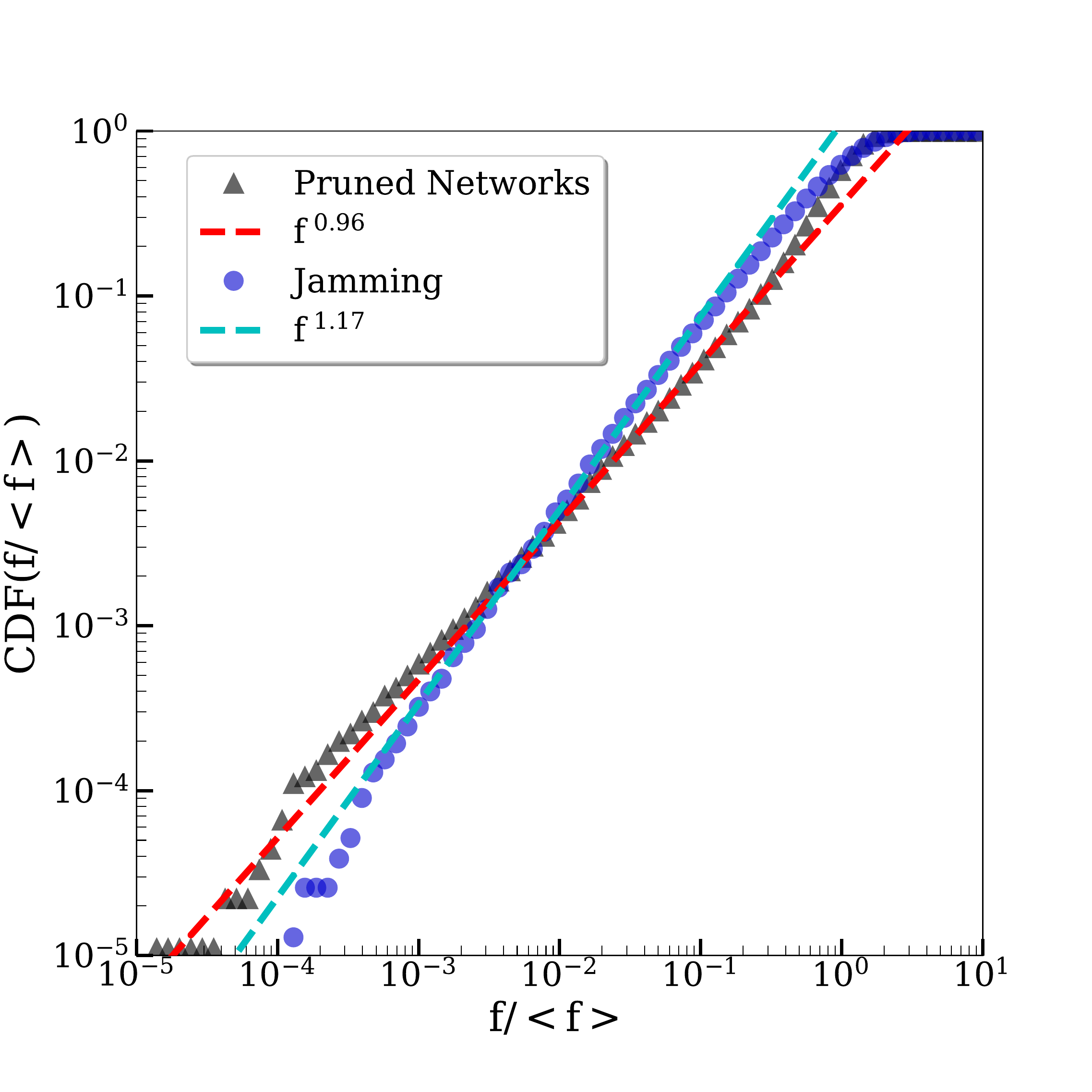}
\caption{(color online) The cumulative distribution function of forces for pruned networks (gray triangles) and jammed systems (blue circles) at isostatic plus one.  Best fit power laws are over plotted in red for the pruned networks and teal for the jammed systems.}
\label{fig:CDForces}
\end{figure}

Figure~\ref{fig:PDForces} shows the probability distribution of forces at isostatic plus one for the pruned networks and the jammed systems.  While they look quite similar on this scale, a plot of the cumulative distribution of forces (Figure~\ref{fig:CDForces}) reveals an intriguing distinction. The physically jammed packing has a low force scaling exponent for all forces that is consistent with the mean field full-replica symmetry breaking results \cite{charbonneau_jamming_2015}, as is expected for a jamming transition that happens deep within the marginal glass phase. However, the pruned network has an exponent in the CDF consistent with $1$, which matches well with the single-replica symmetry breaking result for stable glasses\cite{franz2016simplest}.

\end{document}